\documentclass[conference]{IEEEtran}
\IEEEoverridecommandlockouts
\usepackage{cite}
\usepackage[utf8]{inputenc}
\usepackage{amsmath,amssymb,amsfonts}
\usepackage{algorithmic}
\usepackage{graphicx}
\usepackage{textcomp}
\usepackage{booktabs}
\usepackage{makecell}
\usepackage{tabularx}
\DeclareUnicodeCharacter{202C}{\,}
\usepackage{float}
\usepackage{wasysym}
\usepackage{subcaption}
\usepackage[printonlyused, withpage]{acronym}
\usepackage{xcolor}

\usepackage[colorlinks,linkcolor={blue},citecolor={red},urlcolor={red}]{hyperref}
\def\BibTeX{{\rm B\kern-.05em{\sc i\kern-.025em b}\kern-.08em
    T\kern-.1667em\lower.7ex\hbox{E}\kern-.125emX}}

\begin{document}

\title{Reproducible Optical Tracking Precision: Evaluating a Static, Near-Parallel Support Structure for OptiTrack PrimeX22 Cameras\\
}

\author{\IEEEauthorblockN{1\textsuperscript{st} Oliver Krumpek}
\IEEEauthorblockA{\textit{dept. Machine Vision} \\
\textit{Fraunhofer IPK}\\
Pascalstr. 8-9, 10587 Berlin, Germany \\
oliver.krumpek@ipk.fraunhofer.de}
\and
\IEEEauthorblockN{2\textsuperscript{nd} Ole Kroeger}
\IEEEauthorblockA{\textit{dept. Machine Vision} \\
\textit{Fraunhofer IPK}\\
Pascalstr. 8-9, 10587 Berlin, Germany \\
ole.kroeger@ipk.fraunhofer.de}
\and
\IEEEauthorblockN{3\textsuperscript{rd} Sebastian Mohr}
\IEEEauthorblockA{\textit{dept. Machine Vision}} \
\textit{Fraunhofer IPK}\\
Pascalstr. 8-9, 10587 Berlin, Germany \\
sebastian.mohr@ipk.fraunhofer.de}

\maketitle

\begin{abstract}
This paper presents the design and evaluation of a physical support structure for the OptiTrack X22 tracking systems, constructed from carbon fiber-reinforced polymer (CFRP) and Invar steel. These materials were chosen for their low thermal expansion, ensuring geometric stability and rigidity necessary for accurate spatial measurements. The support system is scalable and adaptable for various applications and setups. The study further investigates the effects of camera placement and separation in near-parallel configurations on measurement accuracy and precision. Experimental results show a significant correlation between camera distance and measurement precision—closer camera setups yield higher accuracy. The optimized camera arrangement allowed the prototype to achieve accuracies of ±0.74 mm along the camera's line of sight and ±0.12 mm in orthogonal directions. The experiments show that the standard deviation of the noise on a single measurement plane orthogonal to the camera's line of sight vary between 0.02 and 0.07, indicating that the measurement noise is not constant for every point in space. Details of the system's design and validation are provided to enhance reproducibility and encourage further development in areas like industrial automation and medical device tracking. By delivering a modular solution with validated accuracy, this work aims to promote innovation and practical application in precision tracking technology, facilitating broader adoption and iterative improvements. This approach enhances the accessibility and versatility of high-precision tracking technology, supporting future progress in the field.

\end{abstract}

\section{Introduction}
The correct positioning of workpieces and tools is a key aspect of production technology, significantly influencing efficiency and quality within industrial operations. Optical tracking systems offer a promising solution for controlling and monitoring precise positioning processes. In principle, it is possible to use pre-calibrated systems such as those used in medical technology \cite{ndigitalOpticalNavigation} or flexible systems with customizable camera arrangements\cite{optitrackCameras}. Pre-calibrated systems are often limited for the requirements of industrial applications due to the number of cameras (usually two), the resulting line-of-sight problem and lack of flexibility. For flexible systems such as the OptiTrack system, calibration is required after each repositioning.  This recalibration impacts direct influence on the measurement accuracy leading to offsets in the announced manufacturer specifications and increases the potential for human error  
\begin{figure}[htbp]
\centering
\includegraphics[width=0.5\textwidth]{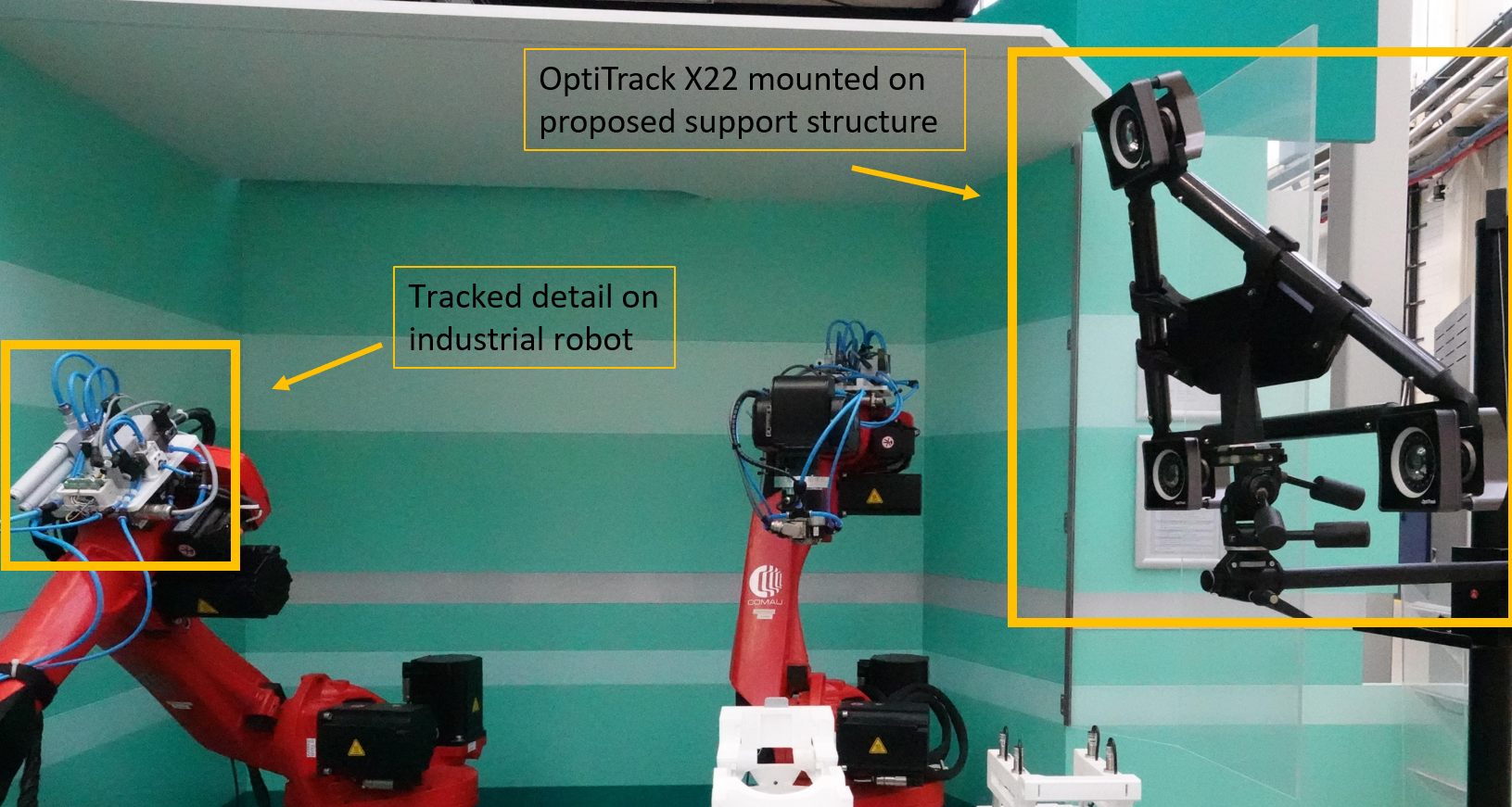}
\caption{Photograph of the presented system in a realistic environment performing measurements on industrial robots }
\label{fig:hardware_setup}
\end{figure}

While manufacturer specifications for measurement accuracy of systems such as OptiTrack X22 Prime are often provided, these values may not directly apply to all individual setups. In the case of the OptiTrack X22 Prime (3D Accuracy of +/- 0.15 mm), the claimed mesurement accuracy based on a setup of 20 cameras distributed across a much larger measurement volume, allowing for tracking from a variety of perspectives, which enhances measurement accuracy but differs from a wide range of possible use cases. Moreover, the mathematical formulation used to calculate the 3D accuracy is not explicitly stated. Our investigation in the selected setup not only evaluates the manufacturer specifications, but also includes experimental investigations to determine the measurement precision and accuracy, with a special focus on the increased measurement error along the camera's line of sight. By exploring these aspects, this paper aims to provide insights into the performance of rigid camera tracking systems and their practical applications in monitoring robotic processes, motion sequences in factory environments, and safety-critical operations, thereby paving the way for further advancements.

\section{Contributions} \label{sec:contributions}

\begin{itemize}
    \item \textbf{Contribution 1:}
    We have developed a static support system for three OptiTrack Prime x22 cameras, which has been optimized in terms of design and material selection for the highest possible rigidity and thus high calibration persistence. The design can be easily adapted to the size requirements of the respective applications by making minimal changes to standard parts. We are making this device publicly available for widespread use, promoting collaboration and further development.\\
    \item \textbf{Contribution 2:} 
    We have determined reliable measurement accuracy values for a static three-camera system with near parallel alignment, which can make a substantial contribution to the design and evaluation of related systems. We clearly point out the relationship between the measurement error and the spatial distribution of targets and cameras. \\
\end{itemize}


\section{Related Work} \label{sec:relatedwork}

In the domain of camera-based tracking, manufacturer specifications typically offer a preliminary indication of system performance. For instance, OptiTrack, a prominent system in this field, asserts an accuracy of ±0.15 mm within a 9 m³ measurement volume \cite{Optitrack.}. However, the achievable accuracy significantly depends on environmental conditions and the number of cameras used. Beyond manufacturer claims, the precision of specific systems or camera configurations has been extensively studied. A 2017 investigation utilizing 21 OptiTrack cameras reported an accuracy of ±0.2 mm across 91\% of a substantially larger 135 m³ volume \cite{Aurand.2017}. Comparable findings emerge from a clinical setting deploying 8 uniformly spaced OptiTrack Flex13 cameras in varied orientations, where the positional error for a single marker was ±0.24 mm, and for a tracked tool, it was as low as ±0.05 mm \cite{Marinetto.2018}. Additionally, research on the errors associated with each axis during translational or rotational movements using seven PrimeX13 cameras revealed errors ranging from 0.001 mm to 0.003 mm translational errors, and from 0.008 mm to 0.025 mm along the camera's viewing axis \cite{PubMedCentral.5152024}. In the context of monitoring handling tasks, a setup comprising four PrimeX41 cameras positioned approximately 1.5m from the target zone measured an error of ±0.22 mm, with a maximum deviation of ±2.30 mm along circular trajectories \cite{Hu.2021}. Zhou et al. (2021) explored the impact of camera layout on the calibration stability and accuracy, obtaining results that align with the calibration outcomes presented in our study \cite{Zhou.2021}. Further investigations into comparable systems by Chen (2021) and Everson (2020) have extended this body of knowledge \cite{Chen.21.08.202122.08.2021}\cite{DanielEverson.2020}. Moreover, research involving a two-camera system highlighted the correlation between measurement distance and detectable noise \cite{MenesesF..2019}.


\begin{figure}[htb]
\centering
\begin{subfigure}{0.24\textwidth}
    \includegraphics[width=\textwidth]{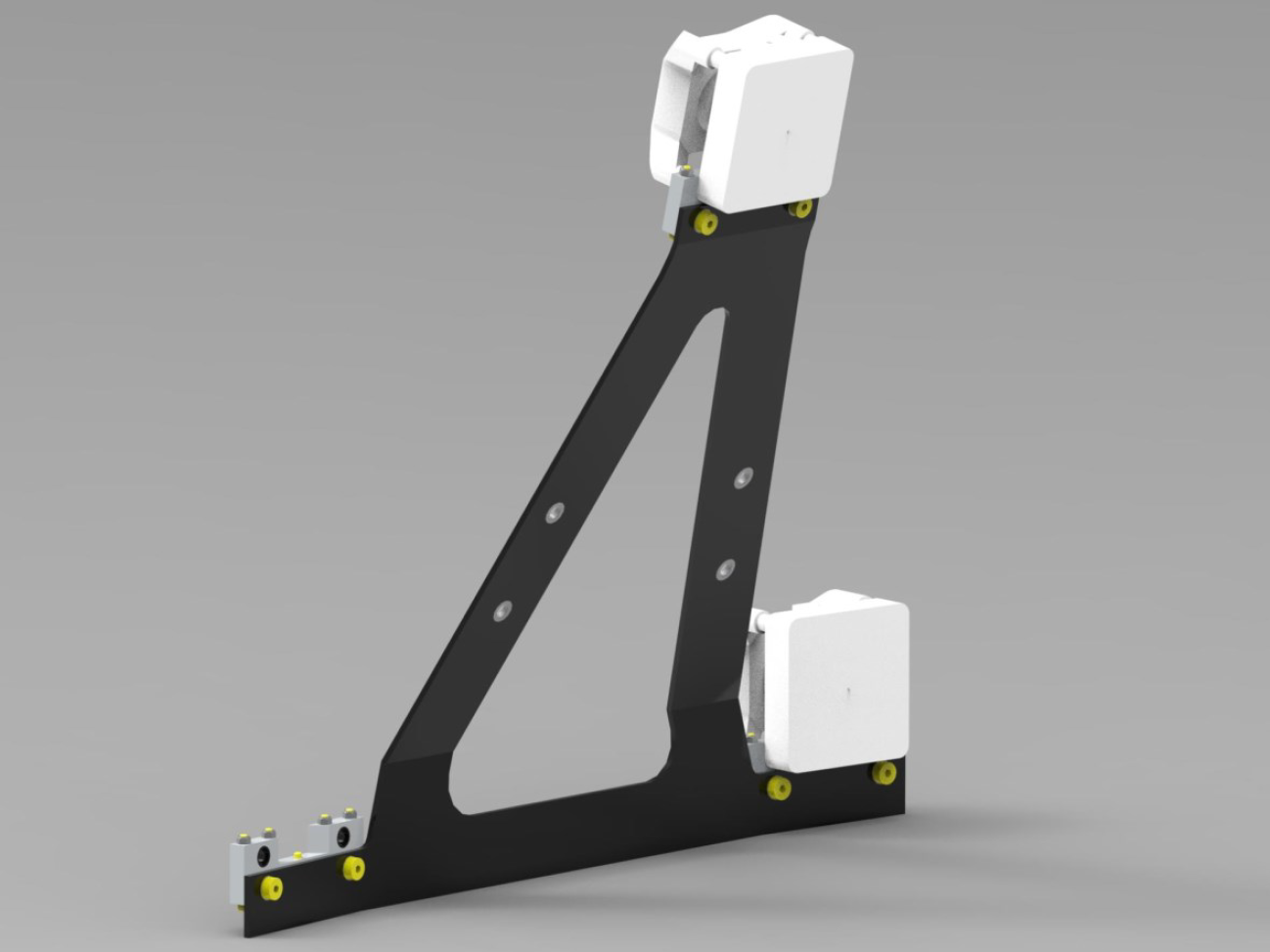}
      \caption{Sketch of the first prototype,  featuring a flat structure design made of carbon fiber.}
\end{subfigure}  
  \begin{subfigure}{0.225\textwidth}
    \includegraphics[width=\textwidth]{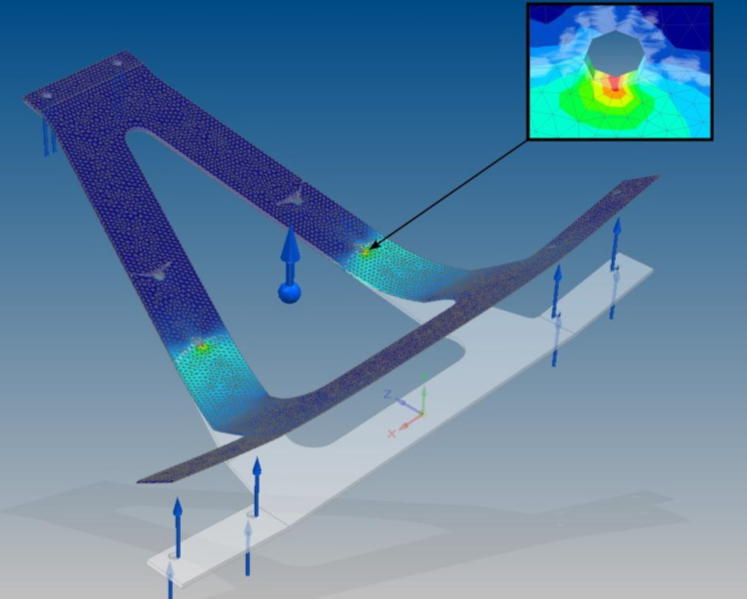}
    \caption{Visualization of potential structural weaknesses under vibrational excitation.}
  \end{subfigure} 
  \caption{Initial development of the support structure, illustrating the foundation for further refinement. This version was found to be costly and prone to structural damage.}
    \label{fig:firstiteration}
  \end{figure}

\section{Design Considerations and Material Selection for Camera Support Structure}

\begin{figure}[b]
\centering
 \includegraphics[width=0.5\textwidth]{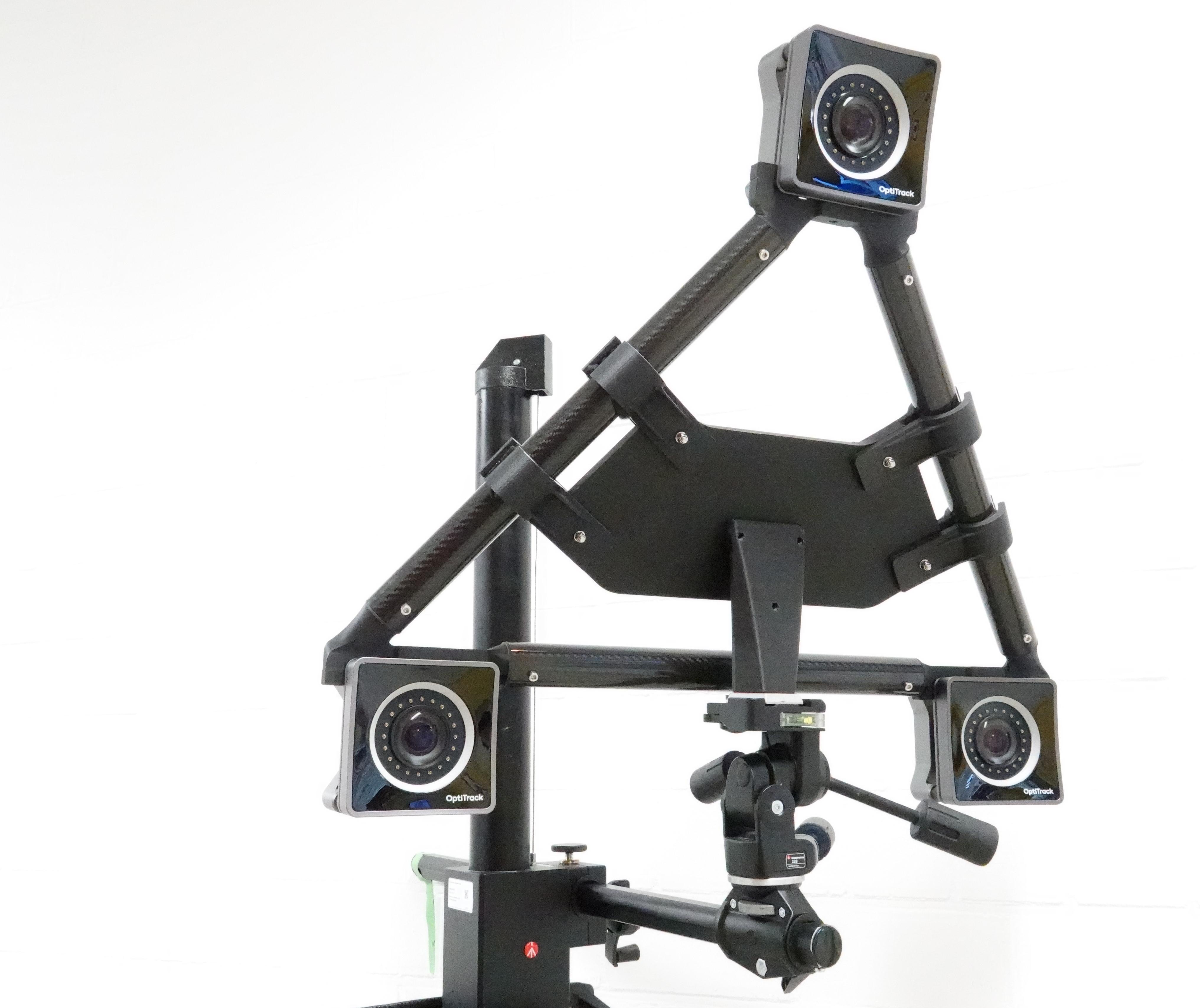}
    \caption{Photo of the final support structure mounted to a camera stand with a Manfrotto camera . A printable version is available on our github repository.}
\label{fig:finalsetup}
\end{figure}

The proposed system is designed for monitoring tool trajectories in robotic systems, machine tools, or additive manufacturing environments, which are often challenging to observe from multiple angles. To address this, a fixed, a compact system with a measurement distance of 200-300 cm was developed. In optical position tracking, both flexible and fixed-camera configurations are viable, with fixed setups eliminating the need for repeated spatial calibration. The introduced system is a permanent, rigidly connected solution that enables straightforward operation by personnel without training for calibration processes. The primary technical challenge in the system's design was achieving a support structure with high dimensional stability to preserve the static spatial relationships of the optical measuring units under prolonged use and mechanical stress. The material selected for the rigid camera connections must exhibit high mechanical strength and minimal thermal expansion.

\begin{figure}[htbp]
\centering
    \includegraphics[width=0.4\textwidth]{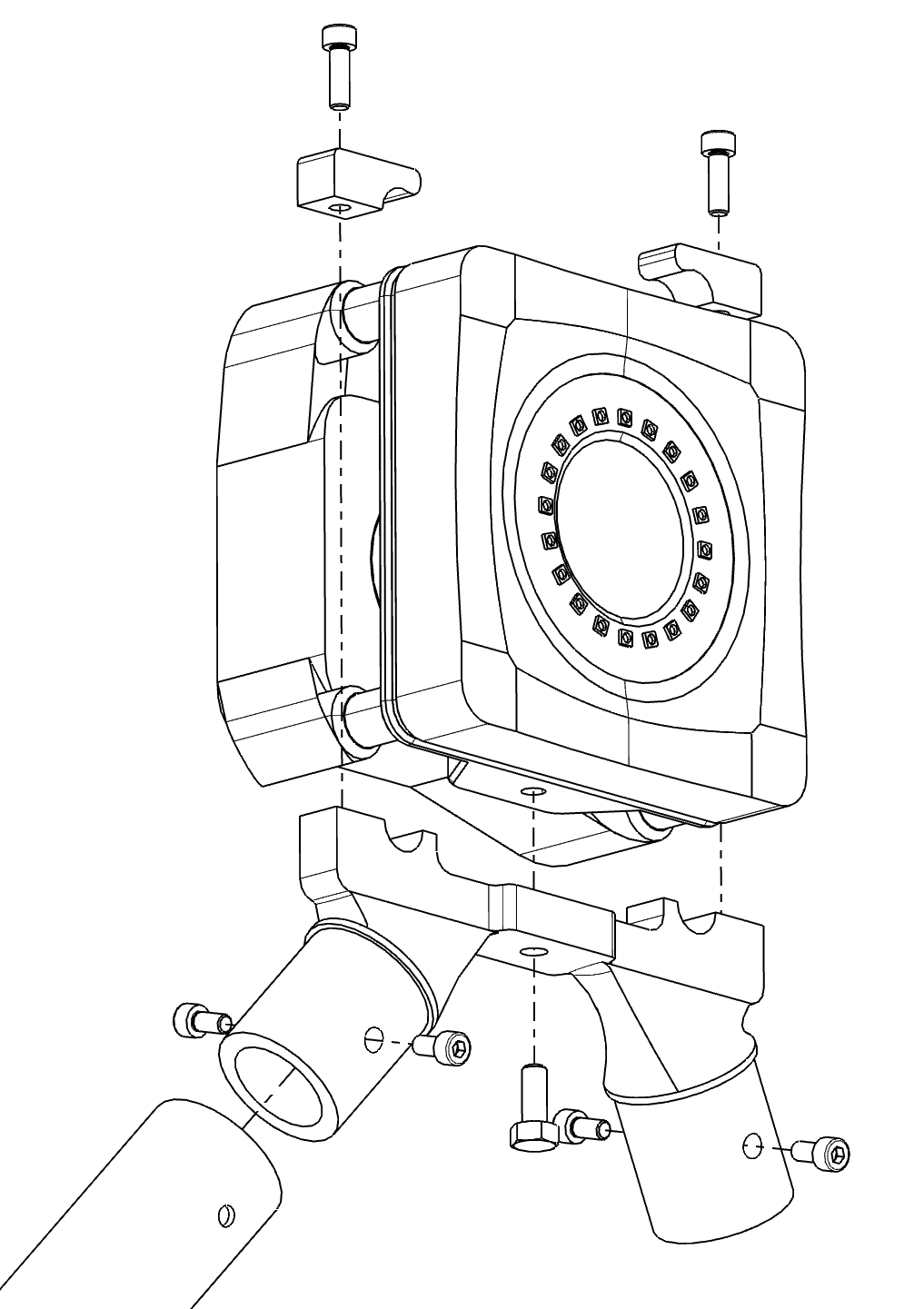}
      \caption{Detailed view of the fastening elements and the screw fittings}
\label{fig:konstruktion}
\end{figure}

\begin{figure}[htbp]
\centering
\includegraphics[width=0.48\textwidth]{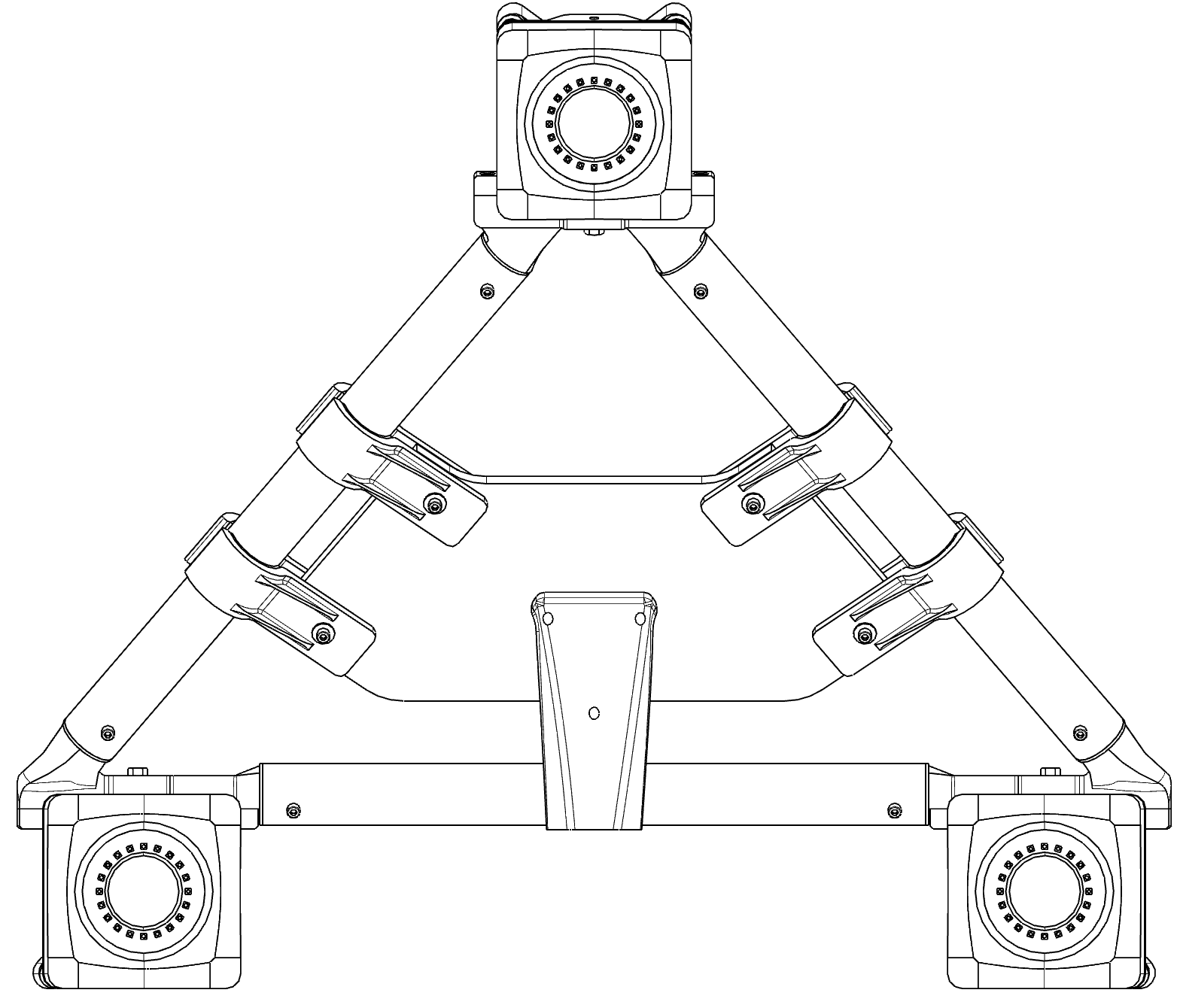}
    \caption{Wireframe of the support structure we presented for the use with three OptiTrack X22}
\label{fig:konstruktion1}
\end{figure}

The support structure in figure \ref{fig:firstiteration} illustrates the first design iteration, comprising a Carbon Fibre Reinforced Polymer (CFRP) laminate, to address the previously outlined requirements. This structure incorporates mounts for a tripod and Invar steel camera brackets, selected for their minimal thermal expansion. The brackets secure the cameras via an integral thread and two clamps, ensuring protection against accidental repositioning. The third camera is omitted from the illustration to highlight the mount design. Despite the benefits of Invar and CFRP, the first iteration encountered issues related to structural damage, inflexibility in modifying the camera setup, and high manufacturing costs, prompting a second development phase.

Informed by the limitations of the first iteration, a modular support system was created using CFRP tubes selected for their adequate stiffness, availability and flexibility. These tubes were connected to the cameras via additively manufactured Invar steel camera adapters (see fig. \ref{fig:konstruktion}). The support structure was clamped onto the pipe structure, thus remaining decoupled. A Finite Element Method (FEM) analysis was carried out to validate the design's robustness under static loads. Forces applied as a result of camera-induced stress at specific mounting points showed maximum stress levels of 2.704 N/mm² in the second load case, with minimal displacements remaining within acceptable limits for calibration adjustments. Extreme repositioning post-calibration should be empirically evaluated for its impact on measurement accuracy. Additionally, a frequency analysis spanning 0 to 1000 Hz identified the critical eigenmode in the final design, revealing peak stresses of 149.85 N/mm² at 661.33 Hz, particularly at the lower camera adapters. These stress values are deemed safe for the system's operational integrity.

\section{Experimental Setup}

  
\subsection{System Calibration}
Camera calibration is essential when setting up and commissioning most optical measurement systems. It involves the mathematical determination of the transformations between the optical centres of the cameras. Calibration is required whenever the geometric relationships between cameras change and is performed using the manufacturer's software, Motive. Calibration accuracy is evaluated using two parameters: Mean Wand Error, which is the average error in the measured length of the calibration wand, and Mean Ray Error, which is the reprojection error and the average distance from the cameras' line of sight to the calculated 3D point. Alexander Schepelmann et.al demonstrated that these values were in line with independently collected evaluation data for a set of 16 cameras \cite{AlexanderSchepelmann.6132022}. The manufacturer categorizes results from poor to exceptional quality. Achieving an ideal calibration is challenging due to inherent numerical approximations and errors. During testing, the goal was to achieve Mean Wand Error and Mean Ray Error within the system's specified accuracy (±0.2 mm). However, out of approximately 40 calibrations, this was achieved in only a few cases, with average values between 0.6 mm for ray- and 0.2 mm for wand-error.\cite{JochemvandenBroek1684922.}\cite{naturalpoint}\cite{Chen.21.08.202122.08.2021}

\begin{figure}[htbp]
\centering
\includegraphics[width=0.48\textwidth]{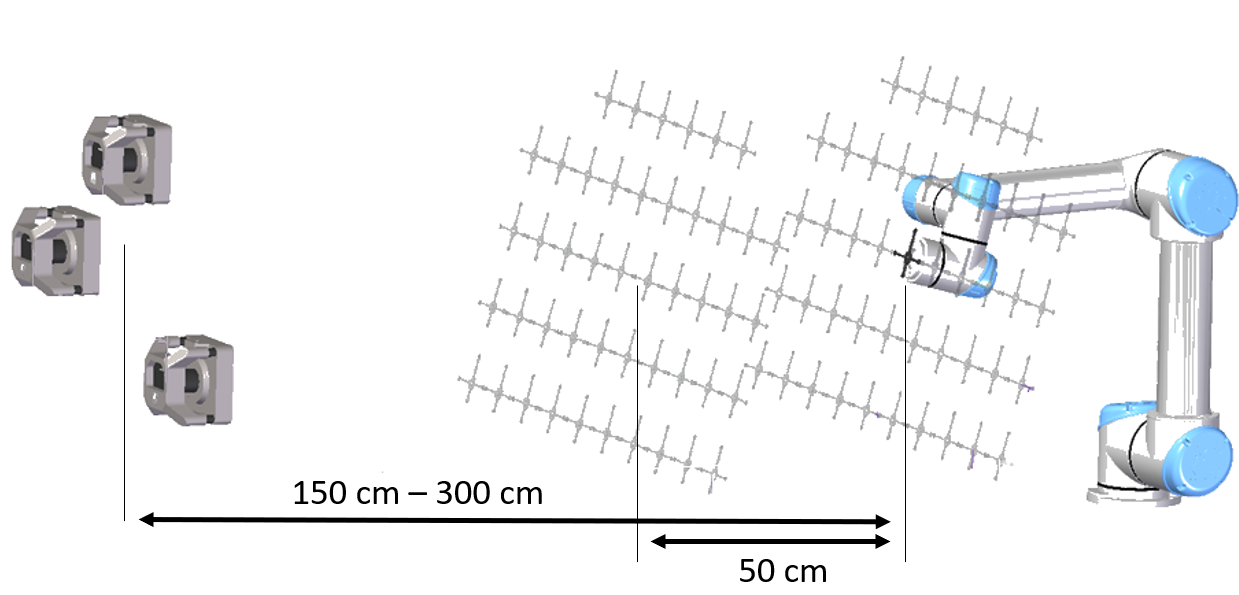}
    \caption{Wireframe of the support structure we presented for the use with three Optitrack X22}
\label{fig:wireframe}
\end{figure}

\subsection{Experiment 1: Investigation in different camera setups}

The first set of experiments was designed to investigate two key relationships: the effect of measurement distance on measurement precision, and the effect of the physical size of the camera setup on measurement precision. Precision was assessed by measuring defined spatial points within a grid pattern on a plane orthogonal to the cameras' line of sight. These measurements were taken at rest to determine the precision at different distances from the cameras, as shown in figure \ref{fig:experiment1}. A total of six distance variations were performed, ranging from a compact 25 cm to an expansive 50 cm, with increments of 5 cm. Similarly, the measurement distance was varied from 150 cm to 350 cm, with increments of 50 cm. To streamline the execution of the experiment, a UR5 robotic arm was used to perform the grid movement with high repeatability accuracy of ± 0,03 mm \cite{universalrobots}. 
\begin{figure}[htbp]
\centering
 \includegraphics[width=0.48\textwidth]{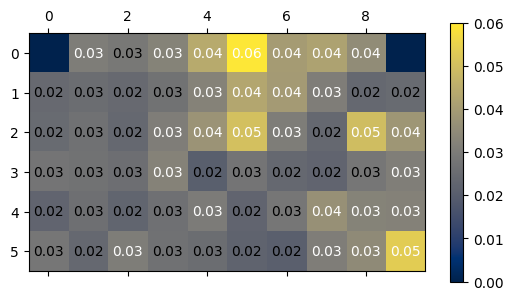}
    \caption{Visualisation of the spacial distribution of the noise on one of the measurement layer orthogonal to theline of sight.}
\label{fig:heat}
\end{figure}

\subsection{Experiment 2: Measurement Accuracy for Linear Displacements}

\begin{figure}[htbp]
\centering
 \includegraphics[width=0.48\textwidth]{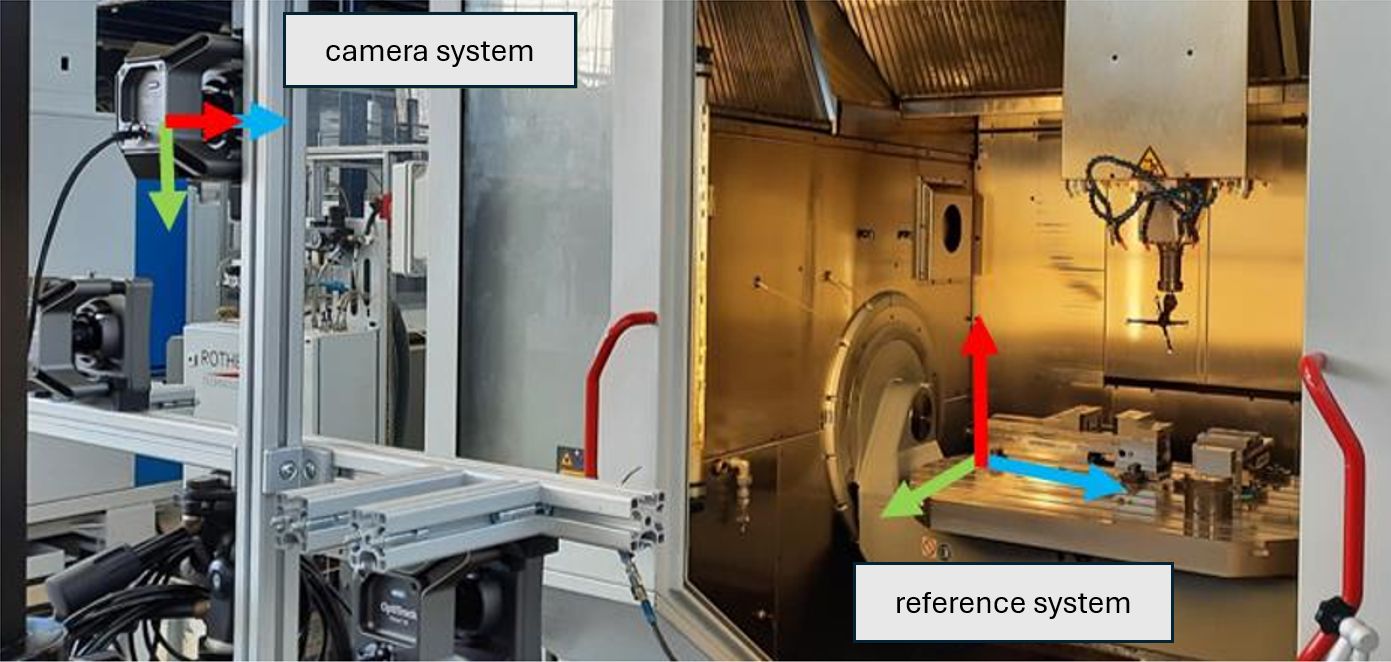}
    \caption{Photo of the navigation camera and the Hermle C50U machine tool, depicting the camera's reference system and the fixed coordinate system (reference/world coordinate system) on the machine tool, which served as the basis for all measurements (left).}
\label{fig:experiment1}
\end{figure}

The aim of this experiment was to determine the measurement accuracy of the camera setup for linearly traversed distances. Unlike the first set of experiments, which focused on positional noise at rest, this series evaluated the measured distance against a known trajectory. The high-precision machine tool Hermle C50U (see fig. \ref{fig:experiment1}), capable of very accurate movements (+/- 0.5 µm), was used to displace the trackers. By moving a tracker through the machine tool in all three axes and capturing the positional data with the camera system, a dataset was created that included both the intended (nominal) and actual (measured) trajectories.

\section{Results}

\begin{table}[htb]
\centering
\begin{tabularx}{0.45\textwidth}{c||c|c}
\toprule
\textbf{Cams.} & $\diameter$ \textbf{ Dist. Error} Slow  & $\diameter$ \textbf{ Dist. Error} Fast (x 1.5) \\
\midrule
2 & 0.89mm & 1.09mm \\ 
3 & 0.33mm & 0.74mm \\ 
5 & 0.33mm & 0.24mm \\ 
\bottomrule
\end{tabularx}
\caption{Average Distance Errors correlated to the number of used cameras for a fast and a slow movement.}
\label{tab:disterror}
\end{table}

\subsection{Observation 1: Increased Measurement Noise with decreased Camera Count and increased Tracker Velocity}

In the experiment we analysed the position of four markers on a tracker attached to the UR5 robotic arm's end-effector, which executed a predefined undulating motion with varied amplitudes and velocities to mimic a waving hand. We investigated the relationship between measurement noise and the number of cameras, using the known Euclidean distances between markers as ground truth. The results indicated that increasing the number of cameras reduces measurement noise. enhancing precision in motion tracking (see Table \ref{tab:disterror}). The maximum error observed was an offset of 1.09 mm.

An additional observation from the data highlights the relationship between tracker velocity and measurement noise. The table also reveals that average deviations are more pronounced during faster motion sequences compared to slower ones. This pattern suggests that the system's ability to accurately track motion diminishes as the speed of the tracked object increases. Addressing this challenge might involve optimizing the camera system's sampling rate, enhancing processing power, or implementing more sophisticated motion prediction algorithms. 

\subsection{Observation 2: Decreased Measurement Precision with Increased Measurement Distance}

\begin{figure}[tb]
\centering
\begin{subfigure}{0.41\textwidth}
    \includegraphics[width=\textwidth]{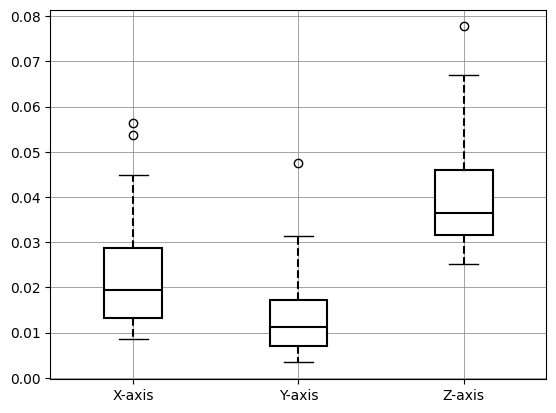}
    \caption{Standard deviation of measurement noise for the proposed hardware setup at a distance of 250 cm.}
\end{subfigure}  
\hspace{0.01\textwidth}
  \begin{subfigure}{0.4\textwidth}
    \includegraphics[width=\textwidth]{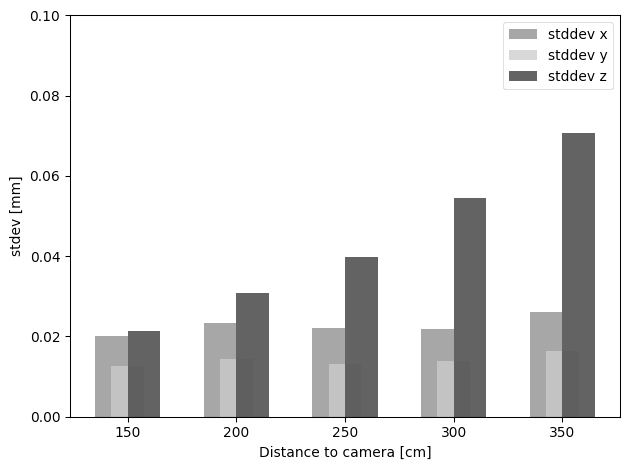}
    \caption{Comparison of standard deviation of measurement noise for increasing measurement distances for the proposed hardware setup.}
  \end{subfigure} 
\caption{Measured standard deviations for the proposed hardware setup}
    \label{fig:staticnoise}
  \end{figure}
  
The experiment also evaluated noise levels in the positional accuracy of a stationary tracker at different distances from the camera system (see fig. \ref{fig:experiment1}). The precision for each axis in the camera coordinate system was measured by the standard deviation in millimetres (see fig. \ref{fig:staticnoise}). The analysis showed that noise increases with distance on all axes, particularly on the z-axis, which corresponds to the camera system's line of sight. The significant rise in z-axis noise at greater distances highlights the difficulties in preserving depth precision, crucial for applications demanding high accuracy in three-dimensional space (see fig. \ref{fig:staticnoise}). 

\subsection{Observation 3: Mitigating Precision Loss at Greater Distances by Increasing Camera Separation}

To further understand the effect of camera system size on measurement accuracy, we evaluated additional setups with varying distances between the cameras. Figure \ref{fig:experiment1} extends the analysis by illustrating the results for setups with camera separations rangin from d = 15 cm to d = 50 cm. The medium setup, with a distance of 30 cm, reduced the noise to about 0.45 mm, while the largest setup, with a distance of 50 cm, further reduced the noise to about 0.25 mm. These results confirm that increasing the camera separation effectively minimises the distance-dependent measurement noise. Our experimental results also show that the noise in the measurement is not the same for every point in space. Fig. \ref{fig:heat} shows the calculated noise of the single measurement position on a single measurement plane (see fig. \ref{fig:experiment1}) The noise varies from 0.02 to 0.06 mm, with parts of the area more prone to higher noise. This pattern continues across all the measurement planes for the camera setups with higher noise in the same areas.

\begin{figure}
\centering
\begin{subfigure}{0.4\textwidth}
    \includegraphics[width=\textwidth]{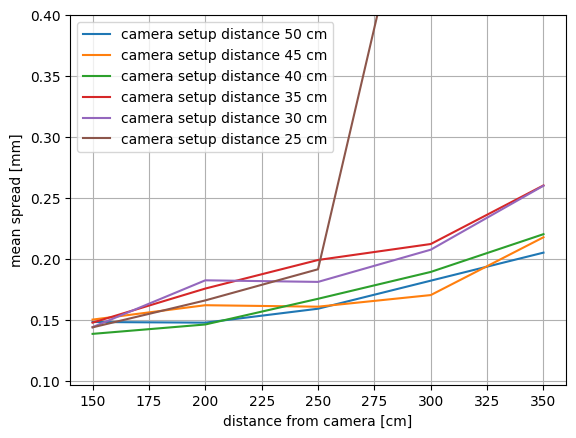}
    \caption{Comparison of standard deviation of measurement noise for different hardware setups.}
\end{subfigure}  
\hspace{0.01\textwidth}
  \begin{subfigure}{0.4\textwidth}
    \includegraphics[width=\textwidth]{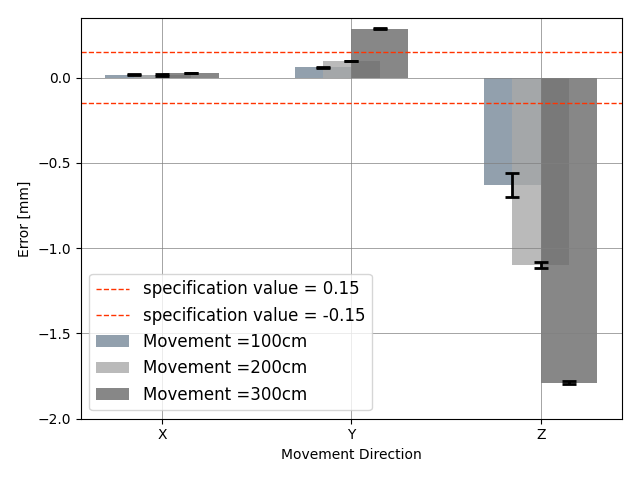}
    \caption{Measured distance error for controlled linear offsets with a precise machine tool, using the proposed hardware setup. }
  \end{subfigure} 
\caption{Experiment results for different camera setups involving an external reference system}
    \label{fig:hermle}
  \end{figure}

\subsection{Observation 4: Heightened Error in Tracking Movements Along the Camera's Line of Sight}

An additional experiment was conducted to assess the system's translational measurement accuracy while tracking a moving object. This experiment compared known distances of linear movements to the distances recorded by the camera system, providing insights into the system's translational accuracy across different axes. Figure \ref{fig:hermle} presents the data for linear displacements, showing the absolute error between the measurement and ground truth values for movements of 100 mm, 200 mm, and 300 mm along the x, y, and z axes. The analysis clearly indicates that inaccuracies were most substantial along the z-axis, which is aligned with the camera's viewing direction. Furthermore, the results demonstrate a consistent relationship between the magnitude of deviations and the length of the movement path, with longer distances yielding greater errors. Notably, for a movement of 300 mm the measurement error on the z-axis was 1.76 mm, compared to around 1.1 mm for 200 mm, and about 0.6 mm for 100 mm.


\section{Discussion}

The primary objective of this study was to design and establish a support structure for a system comprising three OptiTrack Prime X22 cameras, ensuring reliable measurement precision for pre-calibrated setups in challenging environments.
Our experimental results successfully demonstrated the attainable key performance metrics; however, long-term stability considerations were not addressed. The system was constructed with a mount compatible with standard tripods, although specific accessibility needs might require further adaptations.
The experiments carried out focused on measuring noise and the accuracy of translational measurements at distances relevant to the authors. In particular, rotational measurement accuracy was not assessed at this stage.
In the presented configuration with parallel camera alignment, the experiments indicated significant inaccuracies, particularly along the z-axis, which is aligned with the cameras' line of sight. This finding is consistent with the principles of triangulation, where pixel values from camera images of reflective markers yield higher pixel density per unit distance at shorter distances, thereby enhancing system resolution [px/m] with increasing proximity to the camera. The prominent perspective disparities between cameras facilitate improved differentiation of pixel coordinates, further augmenting resolution.
These findings underscore the importance of meticulous consideration regarding camera alignment and distance settings to meet high-precision demands. For optimal performance with the proposed system, it is advisable to keep camera distances below 250 cm; however, further investigation is required  to determine the minimum effective distance.
Noise levels were notably reduced in setups with larger camera separations, as detailed in the Results section. For specific project requirements, minimal adjustments can be made to the presented system, whose design is constrained to cost-effective standard components.

 \newpage

\bibliographystyle{abbrv} 
\bibliography{literature} 

\end{document}